\documentstyle[preprint,prl,aps,epsfig]{revtex}

\tightenlines

\begin{document}
%\twocolumn
\draft
\title{Concentration of electric dipole strength below
the neutron separation energy in N=82 nuclei
}
\author{A.~Zilges, S.~Volz,
M. Babilon~,
T.~Hartmann,
P.~Mohr, and
K.~Vogt}
\address{
  Institut f\"ur Kernphysik, Technische Universit\"at Darmstadt,
  Schlossgartenstrasse 9, D--64289 Darmstadt, Germany
}
\date{\today}
\maketitle
\begin{abstract}
The semi-magic nuclei $^{138}$Ba, $^{140}$Ce, and $^{144}$Sm have
been investigated in photon scattering experiments up to
an excitation energy of about 10 MeV. 
The distribution of the electric dipole strength
shows a resonance like structure 
at energies between 5.5 and 8 MeV
exhausting up to 1\,\% of the isovector E1 Energy Weighted Sum Rule.
\end{abstract}


% 21.10-k Properties of nuclei; nuclear energy levels 
% 23.20.-g Electromagnetic transitions
% 25.20.Dc Photon absorption and scattering
% 27.40.+z  90<_ A <_ 149

%\begin{multicols}{2}
\narrowtext
Collective electric dipole excitations (E1) are common phenomena
in very different finite fermion systems. 
They were first discovered in atomic nuclei where
they can be interpreted as  
out of phase oscillations of protons versus  
neutrons and are characterized as Giant Dipole 
Resonance (GDR) \cite{Bal47,Gol48}. 
In recent years they have also been observed in
metallic clusters where compounds 
of typically 2-40 positively charged ions are oscillating 
versus its electron cloud leading to an E1 resonance like
structure, see e.g.~\cite{Deh87,Iac92,Hab99}. 
Very recently a similar mode has been 
predicted to occur in quantum dots where the fermions
are the electrons and holes confined to a tiny space 
\cite{Del00,Cap01}.

The mean energy of the GDR in nuclei lies around 14 MeV 
in heavy nuclei and around 20 MeV in light nuclei. The GDR exhausts 
about 100\,\% of the Energy Weighted Sum Rule (EWSR) 
for isovector E1 strength.
Microscopically the GDR consists of particle--hole excitations
across one major shell. The residual interaction 
shifts the energy to the GDR region. In the past there
has been some experimental evidence, 
that a considerable part of the E1 strength remains around the 
1$\hbar\omega$\ region
\cite{Iga86,Las86,Ala87,Kop90,Her97,Gov98,Her99}.
This strength is usually specified as
Pygmy Dipole Resonance (PDR). 
However, no systematic study on a number of nuclides
in the energy range up to the particle threshold has been
performed so far. In addition, very different and
sometimes contradictory, collective pictures have 
been proposed to account for this low lying E1 strength.

A nonstatistical distribution of E1 strength close
to the neutron threshold has important implications
for ($\gamma$,n) reaction rates in hot stellar
scenarios important for nucleosynthesis
\cite{Gor98,Moh00}. Please note 
that this is true even for E1 strength below the
threshold because of finite population of
low lying excited states in thermal equilibrium
shifting the collective E1 strength (Brink hypothesis).

Detailed experimental knowledge about E1 excitations below 
the GDR is typically limited to the energy region
below 4.5 MeV. High resolution
photon scattering experiments (Nuclear Resonance
Fluorescence -- NRF) have yielded a wealth of information,
especially for deformed nuclei well away from
shell closures \cite{Kne96,Fra98}. 

In most spherical nuclei only the lowest 
1$^{-}$\ state has been studied. Systematic
($\gamma,\gamma'$) studies together with
complementary experiments with hadronic probes have
established this lowest E1 excitation as a two phonon
quadrupole--octupole $2^{+}\otimes3^{-}$\ mode \cite{Wil98}.
It has a dominant isoscalar character \cite{Poe92}.

For the energy range above the two phonon state
up to the neutron threshold the advent of 
experimental facilities for high precision 
photon scattering at higher energies
enables now a detailed determination of 
the E1 strength pattern.
Photon scattering is an ideal tool to investigate
collective low spin modes because it selectively
populates dipole (and to less extent quadrupole)
excitations from the ground state. The well known
electromagnetic excitation mechanism allows a model
independent determination of absolute transition
strengths (or of dynamic dipole moments). Using a continuous
bremsstrahlung spectrum allows a scan of a wide
energy region with a single measurement \cite{Kne96}.

We have performed photon scattering experiments
on the semi--magic N=82 nuclei
$^{138}$Ba, $^{140}$Ce, and $^{144}$Sm at the
real photon facility of the superconducting 
Darmstadt electron accelerator S--DALINAC which allows
experiments up to E$_{max}$=9.9\,MeV \cite{Moh99}. 
Table I lists the parameters of the experiments: 
Target composition, isotopic enrichment, mass, 
neutron separation energy, maximum energy of the 
bremsstrahlung beam, and the total measuring time.
To distinguish direct decays to the ground state from decays
to excited states one has to perform measurements with
at least two different endpoint energies. For $^{138}$Ba
and $^{140}$Ce data is available from older
($\gamma,\gamma'$) experiments up to 6.7 MeV
\cite{Her97,Her99}. For $^{144}$Sm we performed
an additional experiment with a maximum energy 
of 7.6 MeV.

The raw photon scattering spectrum of $^{138}$Ba recorded
with a Ge(HP) detector positioned at 90$^{\circ}$
with respect to the incoming photon beam is shown
in Fig.~1. It shows a number of lines mainly
stemming from the decay of excited states in $^{138}$Ba 
to the ground state. In addition one can identify
lines from transitions in the photon flux calibration 
target $^{11}$B measured simultaneously. The lines
are sitting on a background of mostly nonresonantly
scattered photons which rises exponentially with decreasing
energy. Already from the raw spectrum the nonstatistical
structure of the excitations in $^{138}$Ba can be guessed:
Besides the strong transition at 4025 keV stemming from the
decay of the $2^{+}\otimes3^{-}$\ mode discussed above,
a concentration of lines is observed between 5.5 and 7 MeV.

A measurement of the angular distribution yielded the spin 
of the excitations. Measurements at only two angles
(90$^{\circ}$\ and 130$^{\circ}$) allow for distinguishing
the two possible $\gamma$\ ray spin cascades starting from
the groundstate, populating a J=1 or 2 state and decaying
back to the groundstate  
($0 \rightarrow 1 \rightarrow 0$\ and
$0 \rightarrow 2 \rightarrow 0$) unambiguously.
A recent parity measurement using a monoenergetic
photon beam has already verified negative parity for all
dipole excitations below 6.7\,MeV  (the higher energy range 
has not been measured so far) in $^{138}$Ba \cite{Pie02}. 
From this result and based on systematic grounds we
assume that all observed dipole excitations have
electric character, i.e.~our summed strength are
an upper limit for the E1 strength.

Figure 2 gives the deduced B(E1)$\uparrow$\ strength 
distribution for the three investigated N=82 isotones.
A very similar pattern is observed in these nuclei:
One isolated E1 excitation at 4026, 3644, and 3226 keV
in $^{138}$Ba, $^{140}$Ce and $^{144}$Sm, respectively,
followed by a resonance like structure of E1 strength
between 5.5 and 8 MeV. Please note that no strength is
observed above approx.~8 MeV. This shows that the
observed states are not just statistical E1 excitations
sitting on the tail of the GDR but represent a fundamental
nuclear structure effect.
A similar concentration of E1 strength has been observed in 
$^{116}$Sn and $^{124}$Sn at slightly higher energies \cite{Gov98}.
In the doubly magic nuclei $^{48}$Ca \cite{Har02}
and $^{208}$Pb \cite{Rye02} the E1 strength is much less
fragmented. However, again a concentration of strength
is found around 1$\hbar\omega$\ in both nuclei.
Folding the B(E1) strength distribution with a Lorentzian 
with a width of $\Gamma$=250 keV shows, that there may
be certain substructures in the resonance region. 
Bumps appear at energies of 5.5, 6.7, and 7.8 MeV in
$^{138}$Ba and at slightly lower values in the other
two nuclides \cite{Vol02}.

Elastic photon scattering using tagged photons between
4.5 and 9 MeV with an energy resolution between
about 40 keV and 100 keV have been performed by
Laszewski on Ba and Ce targets with natural isotopic
composition \cite{Las86}. Whereas our integrated
cross sections for $^{138}$Ba are in agreement with Laszewski's
measurement, the results for $^{140}$Ce differ considerably
in the energy region above 6.5 MeV \cite{Vol02}.

The E1 strength for the lowest lying excitation 
(the two phonon mode) and
the summed strength for the remaining states
(belonging to the resonance mode) up to
the endpoint energy E$_{max}$\ are listed in Table~II. 
One can see that the B(E1) value for the 
two phonon $2^{+}\otimes 3^{-}$ state
is nearly identical for the three nuclei. 
We note that these strengths agree well with the values
measured in the pioneering works by Metzger \cite{Met76}
and Swann \cite{Swa77}.
Recently it has been shown that the dynamic dipole moments determined
experimentally for one class of nuclei
(either semi-magic, non-magic or strongly deformed) scale
with the dipole moment calculated from the dynamic
quadrupole and octupole deformations 
$\beta_{2}$\ and $\beta_{3}$\ \cite{Har00,Bab02}.
The systematics of these B(E1) strengths have 
been discussed in ref.~\cite{And01}.

The remaining E1 strength up to the endpoint energy
of the bremsstrahlung varies much more.
In $^{138}$Ba the weighted mean energy is 
6.5 MeV and 0.78(15)\,\% of the EWSR is exhausted.
The weighted mean energy in $^{140}$Ce and $^{144}$Sm
is 6.4 MeV and 6.0 MeV, respectively. The strengths
exhaust 0.33(5)\,\% and 0.24(4)\,\% of the EWSR in
these two nuclides.

Figure~3 gives the summed B(E1)$\uparrow$\ strengths
above the two phonon excitation up to the neutron
threshold for the three investigated N=82 nuclei. 
From the available data points it is impossible to
decide if there is a systematic decrease of the summed
strength with proton number and/or a minimum around Z=60
as expected from some models (see below).
Therefore photon scattering
experiments on the two remaining stable 
N=82 nuclei $^{136}$Xe and $^{142}$Nd in the same
energy range are mandatory.
Nuclei away from the valley of stability
could be investigated in Coulomb 
scattering in inverse kinematics at radioactive beam 
facilities.

What could be the origin of the observed E1 strength?
One possible explanation would be a breaking of
the symmetry of the proton and neutron distributions
in the nucleus due to the formation of a 
neutron or proton skin. Such a skin may oscillate
versus the core leading to an isovector E1 
mode \cite{Moh71,Suz90,Cha94,Ada96}. 
It would be the analogon to the so called Soft
Dipole Mode (SDM) in exotic light nuclei with 
extreme neutron excess where one or two neutrons
form a halo at great distance to the core. This
SDM has been observed at energies of a few hundred keV
with $B(E1)$\ values in the order of one Weisskopf 
unit \cite{Mil83,Han95}. In some stable nuclei an
enrichment of neutrons or protons in the periphery
is predicted and observed as well: Elastic proton
scattering experiments have shown that the rms
radius of the neutrons is slightly larger than that of the
protons in many neutron rich stable nuclei \cite{Bat89}. In addition
recent antiproton absorption experiments at the LEAR 
facility at CERN have shown, that nuclei with neutron separation
energies of less than about 10 MeV (i.e.~nuclei on the
more neutron rich side of the valley of stability)
possess a neutron rich nuclear periphery \cite{Lub98}.
On the other hand, for certain stable nuclei with a closed
neutron shell like e.g.~$^{144}$Sm a proton rich nuclear
periphery has been observed \cite{Lub98}. 
Calculations in a self consistent HFB theory have yielded
a similar result for N=82 nuclei: 
A neutron skin for the neutron rich nuclei and a proton skin
for the nuclei on the more proton rich side \cite{Dob96}.
Such a skin excitation mode is predicted at
energies between 6 and 10 MeV and with a $B(E1)$\
strength in the order of one percent of the EWSR, 
strongly depending on the $N/Z$\ ratio \cite{Moh71,Suz90}. 
The studies mentioned above would predict a
minimum of the E1 strength resulting from skin excitations
for $^{142}$Nd and a very high value for $^{136}$Xe.
Measurements on the isotopes $^{142}$Nd and $^{136}$Xe 
planned for the near future will therefore allow a more concise 
statement about the origin of the E1 strength.

The so called Toroidal Dipole Resonances (TDR) have
been predicted as a completely different E1 excitation mode 
in the same energy region. Here a vortex collective
motion of the nucleons inside the nucleus leads to
isoscalar transversal E1 excitations \cite{Bas93}.
Recent calculations in the relativistic random phase
approximation predict such a mode to occur in stable
nuclei around 10 MeV \cite{Vre00,Vre02}.
A similar mode has been predicted by Col\`o
et al.~\cite{Col00} at slightly higher energies
and has been observed in $\alpha$\ scattering
experiments by Clark et al.~\cite{Cla01}.
If the resonance between 5.5 and 8 MeV observed in
our experiments is really an isoscalar mode it should be
strongly populated in $\alpha$--scattering experiments
under extreme forward angles which can be performed 
e.g.~at KVI in Groningen and RCNP in Osaka.
The transversal character of the mode could be checked
with electron scattering experiments
with small momentum transfer at the S--DALINAC.

Another possibility of generating E1 strength in 
the energy region below the GDR is a local
breaking of the isospin symmetry due to a
clustering mechanism. Here it is assumed
that in certain heavy nuclei a cluster with a different
neutron to proton ratio than the remaining nucleus may
form. An oscillation of this cluster would lead
to a relatively narrow isovector resonance mode 
which should be concentrated around 6\,MeV 
in heavy nuclei \cite{Iac85,Iac02}.

In conclusion we have performed for the first time
a high resolution study of the electric dipole
strength distribution in $^{138}$Ba, $^{140}$Ce, and
$^{144}$Sm up to the neutron threshold.
A resonance like concentration of strength
has been observed in all three nuclei between
5.5 and 8 MeV. Besides a study of the two remaining
stable N=82 nuclei in ($\gamma,\gamma'$) studies,
complementary experiments with hadronic probes and
electrons are necessary to understand the structure
of the excitations. Measurements with monoenergetic 
polarized photons from laser back scattering 
could help to clarify the parities and detailed decay patterns
of the states \cite{Pie02}.

The authors acknowledge the S--DALINAC group around 
H.-D.\ Gr\"af for the support during the photon scattering
experiments. We thank F.~Iachello for his numerous
important contributions to this manuscript. We thank
P.\ von Brentano, V.\ Yu.\ Ponomarev, A.\ Richter, and
J.\ Wambach for valuable discussions and the
Gesellschaft f\"ur Schwerionenforschung (GSI) for
the loan of the $^{144}$Sm isotope material.
This work was supported by the Deutsche Forschungsgemeinschaft
(contracts Zi\,510/2-1 and FOR 272/2-1).

\newpage
\begin{table}
\begin{center}
\begin{tabular}{cccccc}
Target  
        &  Enrichment
        &  Target mass
        &  S$_{n}$ 
        &  E$_{max}$ 
        &  t$_{meas}$ \\

        &   (\%)    
        &   (mg)        
        &   (MeV)        
        &   (MeV)        
        &     (h)\\
\hline
& & & & & \\
$^{138}$Ba$_{2}$CO$_{3}$ & 99.5 &  3011 &  8.6 & 9.2 & 40\\
$^{140}$CeO$_{2}$        & 88.5 &  3160 &  9.2 & 9.9 & 40\\
$^{144}$Sm$_{2}$O$_{3}$  & 96.5 &  726 & 10.5 & 9.9 & 80\\
                         &      &      &       & 7.6 & 40 \\
\end{tabular}
\end{center}
\caption{
Properties of the examined targets and parameters of
the photon scattering experiments: Target, isotopic
enrichments, target mass, neutron separation energy S$_{n}$,
maximum bremsstrahlung energy E$_{max}$, and
measuring time t$_{meas}$.
}
\end{table}
\begin{table}
\begin{center}
\begin{tabular}{cccccc}
Nucleus  
        &  E$_{twph}$
        &  B(E1)$\uparrow_{twph}$
        &  $\overline{E_{res}}$
        &  $\sum$B(E1)$\uparrow_{res}$
        &  EWSR\\

        &   (MeV)    
        &   (10$^{-3}e^{2}fm^{2}$)        
        &   (MeV)        
        &   (10$^{-3}e^{2}fm^{2}$)        
        &     (\%) \\
\hline
& & & & &\\
$^{138}$Ba   &  4.026 & 17(3) & 6.5(13) & 579(111) & 0.78(15)\\
$^{140}$Ce   &  3.644 & 19(2) & 6.4(10) & 247(40)  & 0.33(5)\\
$^{144}$Sm   &  3.226 & 19(3) & 6.0(11) & 201(36)  & 0.24(4)\\
\end{tabular}
\end{center}
\caption{
Energies E$_{twph}$\ and strengths B(E1)$\uparrow_{twph}$
of the two phonon 2$^{+}\otimes3^{-}$\ states,
mean energies $\overline{E_{res}}$\ and
summed E1 strengths $\sum$B(E1)$\uparrow_{res}$\ 
of 1$^{-}$\ states above the
two phonon states up to the neutron threshold, and exhaustions
of the isovector energy weighted sum rule.
}
\end{table}
\begin{figure}
\epsfig{figure=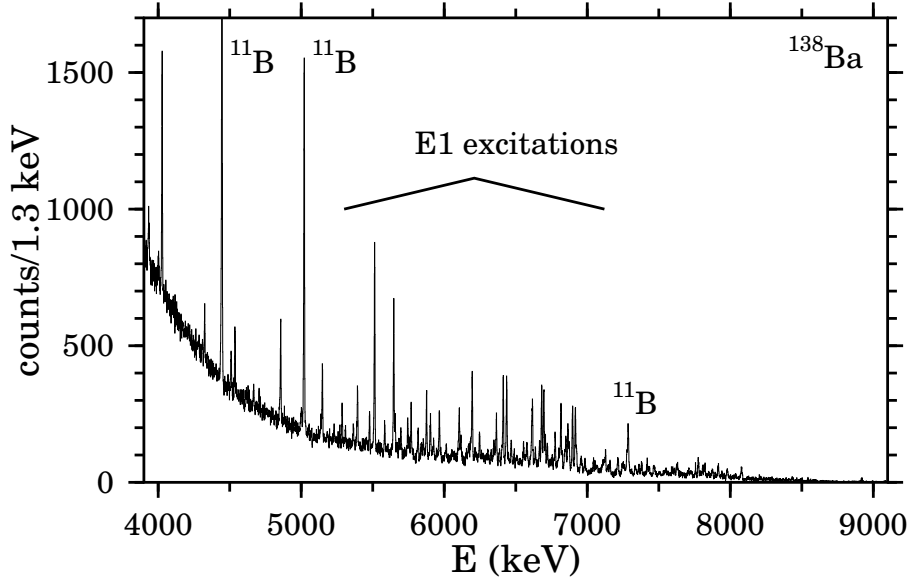,width=12.5cm}
\caption{
Photon scattering spectrum of
$^{138}$Ba measured with an bremsstrahlung endpoint
energy of 9.2\,MeV. Almost all lines stem from
the decay of states populated in $^{138}$Ba.
Lines stemming from the
decay of the $^{11}$B calibration target are labelled.
}
\end{figure}
\begin{figure}
\epsfig{figure=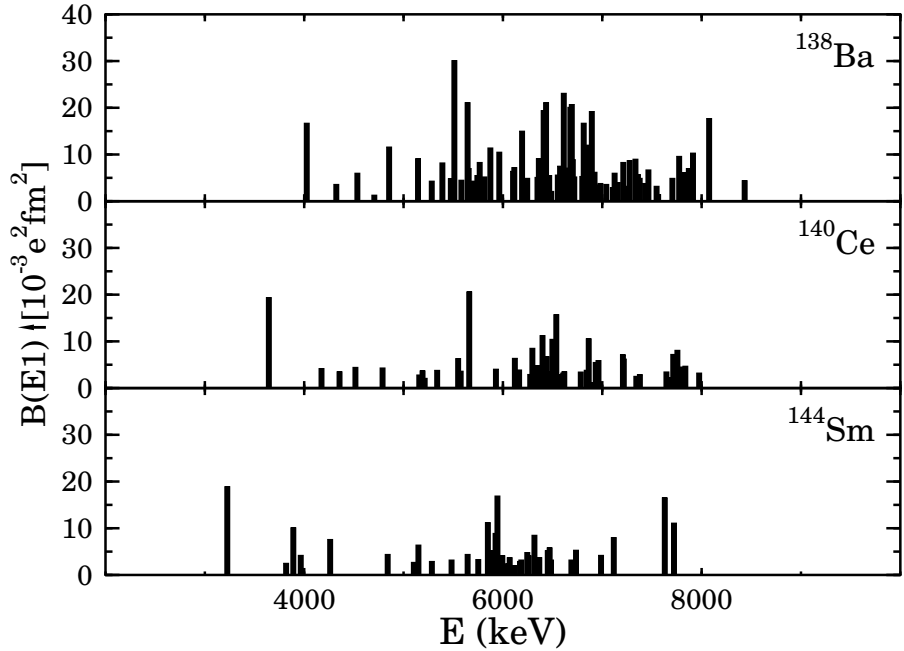,width=12.5cm}
\caption{
Electric dipole B(E1)$\uparrow$\ strength 
distributions below 10 MeV in the three investigated
isotopes $^{138}$Ba, $^{140}$Ce, and $^{144}$Sm.
}
\end{figure}
\begin{figure}
\epsfig{figure=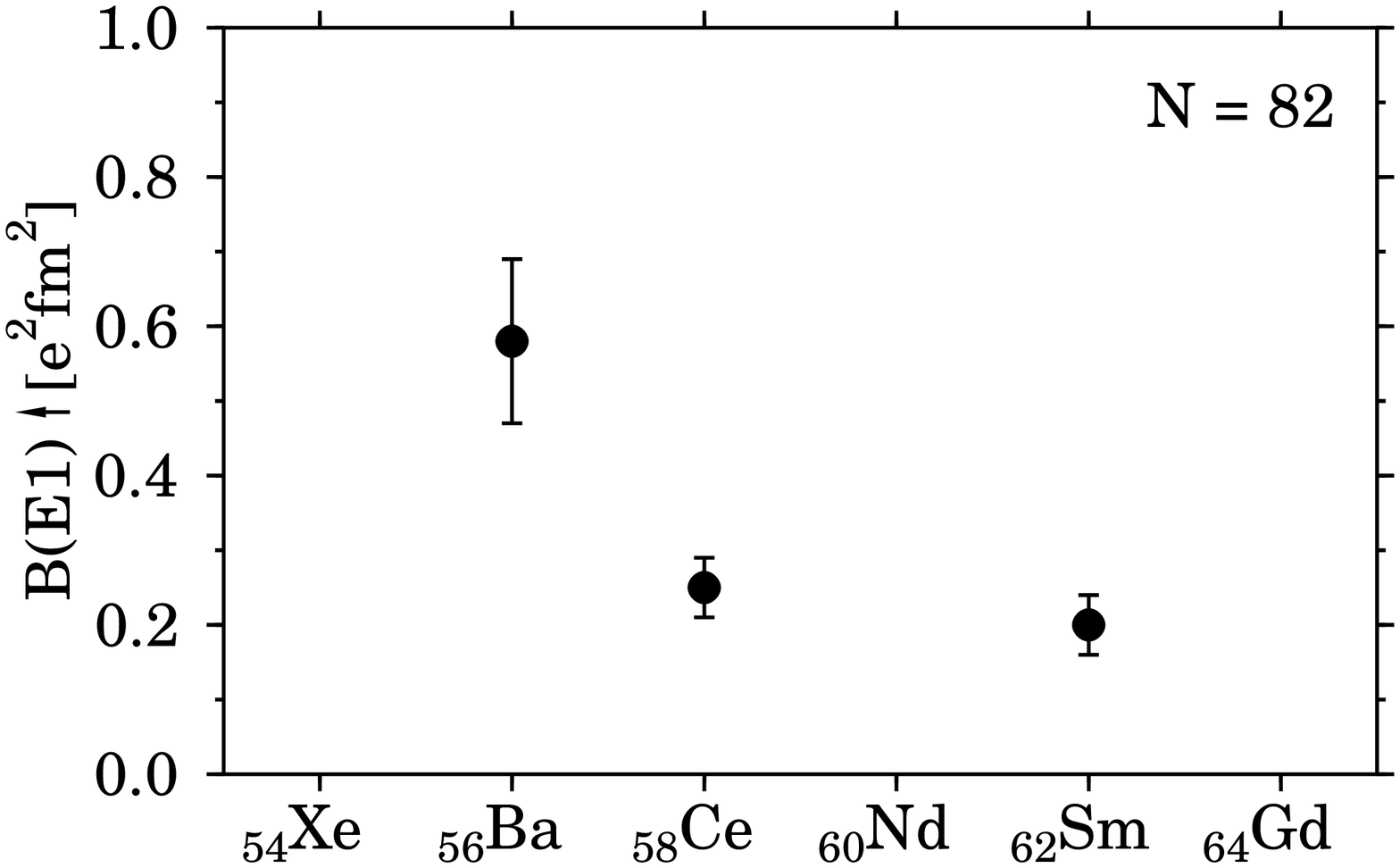,width=12.5cm}
\caption{
Measured summed B(E1)$\uparrow$\ strengths above the
two phonon state up to the neutron
separation threshold. 
}
\end{figure}


\begin{references}
%
\bibitem{Bal47} G.\ C.\ Baldwin and G.\ S.\ Klaiber,
Phys.\ Rev.\ {\bf 71} (1947) 3.
%
\bibitem{Gol48} M.\ Goldhaber and E.\ Teller,
Phys.\ Rev.\ {\bf 74} (1948) 1046.
%
\bibitem{Deh87} 
W.\ A.\ de Heer, K.\ Selby, V.\ Kresin, J.\ Masui,
M.\ Vollmer, A.\ Chatelain, and W.\ D.\ Knight,
Phys.\ Rev.\ Lett.\ {\bf 59} (1987) 1805.
%
\bibitem{Iac92}
F.\ Iachello, E.\ Lipparini, and A.\ Ventura,
Phys.\ Rev.\ {\bf B45} (1992) 4431.
%
\bibitem{Hab99}
H.\ Haberland,
Nucl.\ Phys.\ {\bf A649} (1999) 415c.
%
\bibitem{Del00} 
A.\ Delgado, L.\ Lavin, R.\ Capote, and
A.\ Gonzalez,
Physica\ {\bf E8} (2000) 342.
%
\bibitem{Cap01}
R.\ Capote, A.\ Delgado, and A.\ Gonzalez,
Mod.\ Phys.\ Lett.\ {\bf B15} (2001) 81.
%
\bibitem{Iga86}
M.\ Igashira, H.\ Kitazawa, M.\ Shimizu, H.\ Komano, 
N.\ Yamamuro, 
Nucl.\ Phys.\ {\bf A457} (1986) 301.
%
\bibitem{Las86}
R.M.\ Laszewski,
Phys.\ Rev.\ {\bf C34} (1986) 1114.
%
\bibitem{Ala87}
R.\ Alarcon, R.\ M.\ Laszewski, A.\ M.\ Nathan, and S.\ D.\ Hoblit,
Phys.\ Rev.\ {\bf C36} (1987) 954.
%
\bibitem{Kop90}
J.\ Kopecky and M.\ Uhl,
Phys.\ Rev.\ {\bf C41} (1990) 1941.
%
\bibitem{Her97}
R.-D.\  Herzberg,
%P.\ von Brentano, J.\ Eberth,
%J.\ Enders, R.\ Fischer, N.\ Huxel, T.\ Klemme,
%P.\ von Neumann--Cosel, N.\ Nicolay,
%N.\ Pietralla, V.\ Yu.\ Ponomarev, 
%J.\ Reif, A.\ Richter, C.\ Schlegel,
%R.\ Schwengner, S.\ Skoda, H.\ G.\ Thomas, 
%I.\ Wiedenh\"over, G.\ Winter, and A.\ Zilges,
Phys.\ Lett.\ {\bf B390} (1997) 49.
%
\bibitem{Gov98}
K.\ Govaert, F.\ Bauwens, J.\ Bryssinck, D.\ De\ Frenne,
E.\ Jacobs, W.\ Mondelaers, L.\ Govor, and
V.\ Yu.\ Ponomarev,
Phys.\ Rev.\ {\bf C57} (1998) 2229.
%
\bibitem{Her99}
R.-D.\  Herzberg et al.,
C.\ Fransen, P.\ von Brentano, J.\ Eberth,
J.\ Enders, A.\ Fitzler, L.\ K\"aubler, 
H.\ Kaiser, P.\ von Neumann--Cosel, 
N.\ Pietralla, V.\ Yu.\ Ponomarev, H.\ Prade, A.\ Richter, 
H.\ Schnare, R.\ Schwengner, S.\ Skoda, H.\ G.\ Thomas, H.\ Tiesler, 
D.\ Weisshaar, and I.\ Wiedenh\"over,
Phys.\ Rev.\ {\bf C60} (1999) 051307.
%
\bibitem{Gor98}
S.\ Goriely, 
Phys.\ Lett.\ {\bf B436} (1998) 10.
%
\bibitem{Moh00}
P.~Mohr, K.~Vogt, M.~Babilon, J.~Enders, T.~Hartmann,
C.~Hutter, T.~Rauscher, S.~Volz, and A.~Zilges,
Phys.\ Lett.\ {\bf B488} (2000) 127.
%
\bibitem{Kne96}
U.\ Kneissl, H.\ H.\ Pitz, and A.\ Zilges,
Prog.\ Part.\ Nucl.\ Phys.\ {\bf 37} (1996) 349.
%
\bibitem{Fra98}
C.\ Fransen, O.\ Beck, P.\ von\ Brentano,
T.\ Eckert, R.--D.\ Herzberg, U.\ Kneissl,
H.\ Maser, A.\ Nord, N.\ Pietralla, H.H.\ Pitz,
and A.\ Zilges,\\
Phys.\ Rev.\ {\bf C57} (1998) 129.
%
\bibitem{Wil98}
M.\ Wilhelm, S.\ Kasemann, G.\ Pascovici, E.\ Radermacher,
P.\ von Brentano, and A.\ Zilges,
Phys.\ Rev.\ {\bf C57} (1998) 577.
%
\bibitem{Poe92}
T.\ D.\ Poelhekken, S.\ K.\ B.\ Hesmondhalgh, H.\ J.\ Hofmann, 
A.\ van der Woude, and M.\ N.\ Harakeh,
Phys.\ Lett.\ {\bf B278} (1992) 423.
%
\bibitem{Moh99}
P.\ Mohr, J.\ Enders, T.\ Hartmann, H.\ Kaiser, D.\ Schiesser,
S.\ Schmitt, S.\ Volz, F.\ Wissel, and A.\ Zilges,
Nucl.\ Instr.\ Meth.\ Phys.\ Res.\ {\bf A423} (1999) 480.
%
\bibitem{Pie02}
N.\ Pietralla, Z.\ Berant, V.\ N.\ Litvinenko, S.\ Hartman,
F.\ F.\ Mikhailov, I.\ V.\ Pinayev, G.\ Swift,
M.\ W.\ Ahmed, J.\ H.\ Kelley, S.\ O.\ Nelson, R.\ Prior,
K.\ Sabourov, A.\ P.\ Tonchev, and H.\ R.\ Weller,
Phys.\ Rev.\ Lett.\ {\bf 88} (2002) 012502.
%
\bibitem{Har02}
T.\ Hartmann, J.\ Enders, P.\ Mohr, K.\ Vogt,
S.\ Volz, and A.\ Zilges,
Phys.\ Rev.\ {\bf C65} (2002) 034301.
%
\bibitem{Rye02}
N.\ Ryezayeva, A.\ Richter, private communication (2002).
%
\bibitem{Vol02}
S.\ Volz et al., to be published.
%
\bibitem{Met76}
F.\ R.\ Metzger,
Phys.\ Rev.\ {\bf C14} (1976) 543 and references therein.
%
\bibitem{Swa77}
C.\ P.\ Swann,
Phys.\ Rev.\ {\bf C15} (1977) 1967.
%
\bibitem{Har00}
T.\ Hartmann, J.\ Enders, P.\ Mohr, K.\ Vogt,
S.\ Volz, and A.\ Zilges,
Phys.\ Rev.\ Lett.\ {\bf 85} (2000) 274.
%
\bibitem{Bab02}
M.\ Babilon, T.\ Hartmann, P.\ Mohr, 
K. Vogt, S.\ Volz, and A.\ Zilges,
Phys.\ Rev.\ {\bf C65} (2002) 037303.
%
\bibitem{And01}
W.\ Andrejtscheff, C.\ Kohstall, P.\ von Brentano, 
C.\ Fransen, U.\ Kneissl, N.\ Pietralla, H.H.\ Pitz,
Phys.\ Lett.\ {\bf B506} (2001) 239.
%
\bibitem{Moh71}
R.\ Mohan, M.\ Danos, and L.\ C.\ Biedenharn,
Phys.\ Rev.\ {\bf C3} (1971) 1740.
%
\bibitem{Suz90}
Y.\ Suzuki, K.\ Ikeda, and H.\ Sato,
Prog.\ Theor.\ Phys.\ {\bf 83} (1990) 180.
%
\bibitem{Cha94}
J.\ Chambers, E.\ Zaremba, J.\ P.\ Adams, and B.\ Castel,
Phys.\ Rev.\ {\bf C50} (1994) R2671.
%
\bibitem{Ada96}
J.\ P.\ Adams, B.\ Castel, and H.\ Sagawa,
Phys.\ Rev.\ {\bf C53} (1996) 1016.
% 
\bibitem{Mil83}
D.\ J.\ Millener, J.\ W.\ Olness, E.\ K.\ Warburton, and
S.\ S.\ Hanna, 
Phys.\ Rev.\ {\bf C28} (1983) 497.
%
\bibitem{Han95}
P.\ G.\ Hansen and A.\ S.\ Jensen,
Annu.\ Rev.\ Nucl.\ Part.\ Sci.\ {\bf45} (1995) 591.
%
\bibitem{Bat89}
C.\ Batty, E.\ Friedman, H.\ J.\ Gils, and
H.\ Rebel, 
Adv.\ Nucl.\ Phys.\ {\bf 19} (1989) 1.
%
\bibitem{Lub98}
P.\ Lubinski, J.\ Jastrzebski, A.\ Trzcinska, W.\ Kurcewicz,
F.\ J.\ Hartmann, W.\ Schmid, T.\ von Egidy,
R.\ Smolanczuk, and S.\ Wycech,
Phys.\ Rev.\ {\bf C57} (1998) 2962
% 
\bibitem{Dob96}
J.\ Dobaczewski, W.\ Nazarewicz, and T.\ R.\ Werner,
Z.\ Phys.\ {\bf A354} (1996) 27.
%
\bibitem{Bas93}
S.\ I.\ Bastrukov, S.\ Misicu, and A.\ V.\ Sushkov,
Nucl.\ Phys.\ {\bf A562} (1993) 191.
%
\bibitem{Vre00}
D.\ Vretenar, A.\ Wandelt, and P.\ Ring,
Phys.\ Lett.\ {\bf B487} (2000) 334.
%
\bibitem{Vre02}
D.\ Vretenar, N.\ Paar, P.\ Ring, and T.\ Niksic,
Phys.\ Rev.\ {\bf C65} (2002) 021301.
%
\bibitem{Col00}
G.~Col\`o, N.~Van Giai, P.~F.~Bortignon, and M.~R.~Quaglia,
Phys.\ Lett.\ {\bf B485} (2000) 362.
%
\bibitem{Cla01}
H.~L.~Clark, Y.-W.~Lui, and D.~H.~Youngblood,
Phys.\ Rev.\ {\bf C63} (2001) 031301(R).
%
\bibitem{Iac85}
F.\ Iachello,
Phys.\ Lett.\ {\bf B160} (1985) 1.
%
\bibitem{Iac02}
F.\ Iachello,
private communication (2002)

\end{references}
\end{document}